\documentclass[iop]{emulateapj}
\usepackage{natbib}

\shorttitle{Four Newly Discovered Ultra Metal-Poor stars}
\shortauthors{Hansen et al.}
\submitted{Accepted for publication in the Astrophysical Journal}

\begin{document}

\title{EXPLORING THE ORIGIN OF LITHIUM, CARBON, STRONTIUM AND BARIUM WITH FOUR
  NEW ULTRA METAL-POOR STARS\footnotemark[1]}

\footnotetext[1]{Based on observations made with the
    European Southern Observatory (ESO) Telescopes.}

\author{T. Hansen,\altaffilmark{2} C. J. Hansen, \altaffilmark{2}
  N. Christlieb,\altaffilmark{2} D. Yong,\altaffilmark{3}
  M. S. Bessell,\altaffilmark{3} A. E. Garc\'{i}a P\'{e}rez,\altaffilmark{4}
  T. C. Beers,\altaffilmark{5},\altaffilmark{6} V. M. Placco,\altaffilmark{7}
  A. Frebel,\altaffilmark{7} J. E. Norris\altaffilmark{3} and M. Asplund\altaffilmark{3}}

\altaffiltext{2}{Landessternwarte, ZAH, K{\"o}nigstuhl 12, D-69117 Heidelberg,
  Germany;thansen@lsw.uni-heidelberg.de, cjhansen@lsw.uni-heidelberg.de,
  nchristlieb@lsw.uni-heidelberg.de} 
\altaffiltext{3}{Research School of Astronomy and Astrophysics, The Australian
  National University, Weston, ACT 2611, Australia; yong@mso.anu.edu.au,
  bessell@mso.anu.edu.au,john.norris@anu.edu.au, martin.asplund@anu.edu.au} 
\altaffiltext{4}{Department of Astronomy, University of Virginia,
  Charlottesville, VA 22904-4325, USA; aeg4x@virginia.edu} 
\altaffiltext{5}{National Optical Astronomy Observatory, Tucson, AZ 85719, USA}
\altaffiltext{6}{JINA: The Joint Institute for Nuclear Astrophysics, 225
  Nieuwland Science Hall, Department of Physics, University of Notre Dame,
  Notre Dame, IN 46556-5670, USA; beers@noao.edu}
\altaffiltext{7}{Gemini Observatory, Hilo, HI 96720, USA; vplacco@gemini.edu}
\altaffiltext{8}{Department of Physics, Massachusetts Institute of Technology,
  Cambridge, MA 02139, USA; afrebel@mit.edu}

\begin{abstract}

We present an elemental abundance analysis for four newly discovered
ultra metal-poor stars from the Hamburg/ESO survey, with
$\mathrm{[Fe/H]}\leq-4$. Based on high-resolution, high signal-to-noise
spectra, we derive abundances for 17 elements in the range from Li to
Ba. Three of the four stars exhibit moderate to large over-abundances of
carbon, but have no enhancements in their neutron-capture elements. The
most metal-poor star in the sample, HE~0233$-$0343  ($\mathrm{[Fe/H]} = -4.68$),
is a subgiant with a carbon enhancement of $\mathrm{[C/Fe]}= +3.5$, slightly
above the carbon-enhancement plateau suggested by Spite et al. No carbon
is detected in the spectrum of the fourth star, but the quality of its
spectrum only allows for the determination of an upper limit on the
carbon abundance ratio of $\mathrm{[C/Fe]} < +1.7$. We
detect lithium in the spectra of two of the carbon-enhanced stars,
including HE~0233$-$0343. Both stars with Li detections are Li-depleted, with
respect to the Li plateau for metal-poor dwarfs found by Spite \& Spite. This
suggests that whatever site(s) produced C either do not completely destroy
lithium, or that Li has been astrated by early-generation stars and mixed with
primordial Li in the gas that formed the stars observed at present. The
derived abundances for the $\alpha$-elements and iron-peak elements of the
four stars are similar to those found in previous large samples of extremely
and ultra metal-poor stars. Finally, a large spread is found in the abundances
of Sr and Ba for these stars, possibly influenced by enrichment from fast
rotating stars in the early universe. 

\end{abstract}



\section{Introduction}
The stellar atmospheres of the first generations of low-mass ($M\leq$
0.8 $M_\odot$) stars are expected to retain, to a large extent, detailed
information on the chemical composition of the nearly pristine gas of
the interstellar medium (ISM) at the time and place of their birth.
Detailed abundance analyses of metal-poor stars thus enable studies of
the formation and evolution of the elements in the early Galaxy. At
these early times, the two light elements carbon and lithium play a major
role in cosmological studies, as well as in our understanding of early
star formation. In addition, the production site(s) of the elements
beyond the iron peak remains a major unanswered question.

Recent studies, such as \citet{carollo2012}, \citet{lee2013}, and
\citet{norris2013b} confirm that carbon-enhanced metal-poor (CEMP)
stars\footnote{Originally defined by \citet{beerschristlieb2005} as metal-poor
($\mathrm{[Fe/H]} \leq -1.0$) stars with $\mathrm{[C/Fe]}\geq +1.0$; a level
of carbon enrichment $\mathrm{[C/Fe]} \geq +0.7$ is used in most contemporary
work.} constitute a large fraction of the most metal-poor stars known and that
the fraction of CEMP stars increases dramatically with decreasing
metallicity, accounting for $\sim$40\% of all stars with
$\mathrm{[Fe/H]} \leq -3.5$. In fact, four of the five stars previously known
with $\mathrm{[Fe/H]}< -4.5$ are confirmed CEMP stars
\citep{christlieb2002,frebel2005, norris2007, caffau2011, keller2014}. In this
paper, we add another confirmed CEMP star with $\mathrm{[Fe/H]} < -4.5$, as
described below. 

\citet{beerschristlieb2005} specify a nomenclature that identifies a
number of subclasses for CEMP stars. The CEMP-$s$ stars exhibit
over-abundances of elements predominantly produced by the so-called slow
neutron-capture process, or $s$ process, such as barium. These stars are
the most commonly observed subclass of CEMP stars; around 80\% of CEMP
stars exhibit $s$-process-element enhancements \citep{aoki2007},
including both the CEMP-$s$ and CEMP-$r/s$ subclass (stars that, in
addition to exhibiting $s$-process element enhancement, are also
enhanced in elements predominantly produced in the rapid neutron-capture
process, or $r$ process, such as europium). The favored scenario for the
production of CEMP-$s$ (and CEMP-$r/s$) stars is mass transfer of
carbon- and $s$-process-enhanced material from the envelope of an
asymptotic giant branch (AGB) star to its (presently observed) binary
companion \citep[e.g., ][]{herwig2005,sneden2008}. Observational
evidence now exists to suggest that the CEMP-$r/s$ stars (and other
$r$-process-element-rich stars) were enhanced in $r$-process elements in
their natal gas clouds by previous generations of supernovae (SNe), and
did not require a contribution of $r$-process elements from a binary
companion \citep[see ][]{thansen2011}.

The CEMP-no subclass includes CEMP stars that exhibit no enhancements in
their neutron-capture elements. It has been shown that at extremely low
metallicity, $\mathrm{[Fe/H]} < -3.0$, the CEMP-no stars are the
dominant subclass \citep{aoki2010,norris2013b}. Different progenitors
have been suggested for the CEMP-no stars, such as pollution by faint SNe that
experienced  extensive mixing and fallback during their explosions
\citep{umeda2003,umeda2005,tominaga2007,tominaga2013,ito2009,ito2013,nomoto2013},\footnote{This
  model well-reproduces the observed elemental-abundance pattern of the
  CEMP-no star BD+44$^{\circ}$493, the ninth-magnitude, $\mathrm{[Fe/H]} =
  -3.8$ star (with $\mathrm{[C/Fe]} = +1.3$, $\mathrm{[N/Fe]} = +0.3$,
  $\mathrm{[O/Fe]} = +1.6$) discussed by \citet{ito2009,ito2013}.}
winds from massive, rapidly rotating, mega metal-poor ($\mathrm{[Fe/H]} <
-6.0$) stars, sometimes referred to as ``spinstars''
\citep{hirschi2006,meynet2006,hirschi2007,meynet2010,cescutti2013}, or mass
transfer from an AGB star companion \citep{suda2004, masseron2010}. This
latter explanation encounters difficulties, however, when confronted
with results from recent radial-velocity monitoring programs. 

Radial-velocity data support the expected differences in the
binary nature of CEMP-$s$ (CEMP-$r/s$) and CEMP-no stars.
\citet{lucatello2005} argued that multiple-epoch observations of
CEMP-$s$ stars are consistent with essentially all CEMP-$s$ (CEMP-$r/s$)
stars being members of binary systems. Although more data are desired for
CEMP-no stars, \citet{thansen2013} report that the fraction of binaries
among stars of this class is no higher than expected for random samples
of very metal-poor giants. \citet{norris2013b} reach similar conclusions
for their limited radial-velocity data for a number of CEMP-no stars.

The measured Li abundances of CEMP stars do not show an obvious
correlation with C at the lowest metallicities, but do exhibit a general
downward trend with declining [Fe/H]. \citet{masseron2012} considered
the CEMP stars with measured Li reported in the literature and added 13
new stars to the list; they highlight the large spread in the measured
Li abundances for CEMP stars. In addition to the production of Li during
big bang nucleosynthesis, Li can also be produced via the Cameron--Fowler
mechanism in AGB stars if $^7$Be, created at the bottom of the
convective envelope, captures an electron \citep{sackmann1992}. If CEMP
stars are the result of mass transfer from an AGB companion, then the Li
abundances in CEMP stars will reflect a combination of (1) Galactic
chemical evolution and (2) Li production/destruction in the AGB
companion. Additional data are necessary to explore and test this
hypothesis in more detail. It should also be recalled that
\citet{piau2006} argued that primordial Li could be astrated by
first-generation stars, objects similar in nature to the massive-star
progenitors suggested for CEMP-no stars. In this view, the fact that Li
abundances for CEMP-no stars are always {\it below} the level of the
Spite Li plateau \citep[see e.g., ][]{masseron2012} can be understood as
the result of various degrees of local mixing between Li-astrated
material ejected from first-generation stars and the surrounding gas
having the primordial level of Li.  

Most elements beyond the iron peak are produced by neutron capture,
either in the $s$ process or the $r$ process \citep[e.g.,
][]{burbidge1957,sneden2008}. The neutron-capture elements strontium and
barium are those that are most easily measured in low-metallicity
stars. At solar metallicity, these elements are produced in the main
$s$ process in AGB stars \citep{busso1999,kappeler2011}, but at low
metallicity, AGB stars may not have had time to sufficiently enrich the
ISM. Hence, the Sr and Ba abundances observed in low-metallicity stars are
presumably produced via the main $r$ process, most likely occurring in the
final stages of the life of massive stars \citep{truran1981,thielemann2011},
or in the weak $s$ process suggested to occur in spinstars
\citep{pignatari2008,cescutti2013}. 

More studies of the lowest metallicity stars are required to gain a
deeper understanding of the nucleosynthesis processes taking place in
the early universe, for both the light and heavy elements, since at
present fewer than 10 stars with $\mathrm{[Fe/H]} \leq -4.2$ have been
analyzed. This paper presents four newly discovered ultra metal-poor (UMP)
stars ($\mathrm{[Fe/H]} < -4.0$), three of which are enhanced in
carbon but not in neutron-capture elements, and are hence classified as
CEMP-no stars. We also detect lithium in the spectra of two of the
stars, one of these being the second most metal-poor star with detected
Li known to date.  

\section{Observations and Data Analysis}

The four stars presented in this paper are part of a larger sample of
metal-poor candidates selected from the Hamburg/ESO survey, followed-up
with medium-resolution spectroscopy on a variety of 2$-$4 m class
telescopes, then observed at high spectral resolution with Very Large
Telescope (VLT)/UVES \citep{dekker2000}. The complete sample will be presented
in Paper~II of this series, along with a detailed description of the
observations, data reduction procedure, parameter determination, and abundance
analysis. Here, only the key points of the techniques employed are listed.

\begin{figure*}
\begin{center}
\includegraphics[angle=0,width=6.8in]{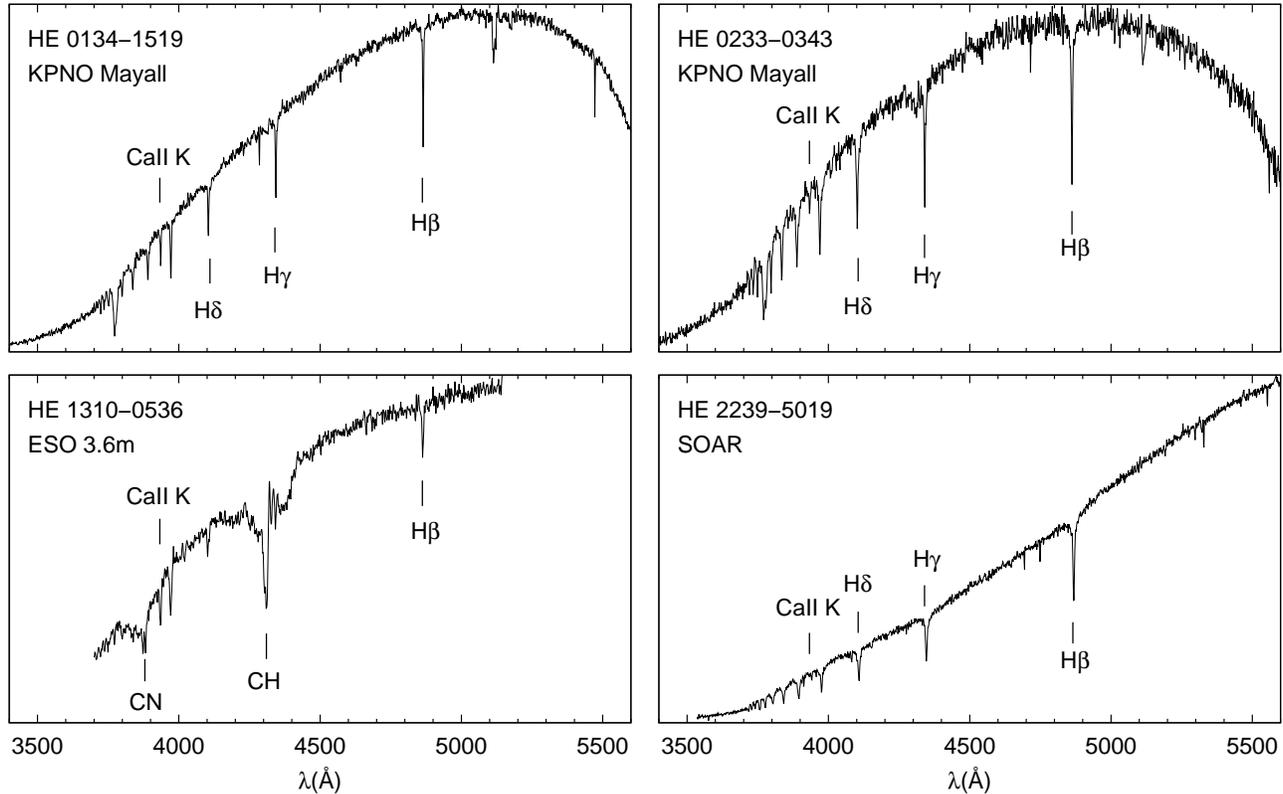}
\end{center}
\caption{Medium-resolution spectra of our four program stars. The locations of
the Ca\,{\sc ii}~K line, $H_{\beta}$, 
$H_{\gamma}$, and $H_{\delta}$ lines are shown. For HE~1310$-$0536, the CH and
CN molecular carbon bands are clearly visible. \label{fig1}}
\end{figure*}

Figure \ref{fig1} shows the medium-resolution spectra of the
program stars. It is possible to see features such as the Ca\,{\sc ii}~K
line, $H_{\beta}$, $H_{\gamma}$, and $H_{\delta}$, as well as the CH and
CN molecular carbon bands for HE~1310$-$0536. Both the Southern Astrophysical
Research (SOAR) 4.1 m and KPNO/Mayall 4 m data have a wavelength coverage of 3550--5500\,{\AA},
with a resolving power of $R\sim 1500$ and signal-to-noise ratios of
S/N$\sim 30$ per pixel at 4000\,{\AA}. For the ESO~3.6 m data, the
resolving power and signal-to-noise were similar to the SOAR 4.1 m and Mayall
4 m data, but the wavelength range is narrower, covering the interval
3700--5100\,{\AA}.

Medium-resolution spectra obtained with the Wide Field Spectrograph
\citep[WiFeS; ][]{dopita2007} on the Australian National University
2.3 m Telescope at Siding Spring Observatory were used for the
temperature determination.

The high-resolution data was obtained during the nights of 2005 November 17
and 20, and 2006 April 17. The data cover a wavelength range from 3100\,{\AA}
to 9500\,{\AA}, with a resolving power of $R \sim$ 45000. The spectra were
reduced using the UVES reduction pipeline, version 4.9.8. Radial-velocity
shifts, co-addition of the spectra, and continuum normalization were all
performed using IRAF\footnote{IRAF is distributed by the National Astronomy
  Observatory, Inc., under cooperative agreement with the National Science
  Foundation.}. The average S/N of the reduced spectra is
S/N $\sim 10$, $\sim 30$, and $\sim 55$ pixel$^{-1}$ at 3400\,{\AA},
4000\,{\AA}, and 6700\,{\AA}, respectively.

\subsection{Stellar Parameters}

The stellar atmospheric parameters were determined by standard
techniques, generally following the steps outlined in \citet{yong2013}.
Effective temperatures were determined by fitting the spectrophotometric
observations with model atmosphere fluxes \citep{bessell2007,
norris2013a}. LTE model atmosphere fluxes from the MARCS grid
\citep{gustafsson2008}, with $\mathrm{[\alpha/Fe]}=+0.4$, were used for
the model fitting. Estimates of surface gravity were determined from the
$Y^2$ isochrones \citep{demarque2004}, assuming an age of 10 Gyr and an
$\alpha$-element enhancement of $\mathrm{[\alpha/Fe]}=+0.3$. These
isochrones only extend down to $\mathrm{[Fe/H]}=-3.5$; therefore, a linear
extrapolation down to $\mathrm{[Fe/H]}=-4.7$ has been used to obtain the
surface-gravity estimates for our four stars. The average difference
between the listed surface gravities, where the actual $\mathrm{[Fe/H]}$
values have been used, and the surface gravity obtained using the
$\mathrm{[Fe/H]}=-3.5$ isochrone, is rather small (on the order of 0.07
dex). Metallicities were determined from equivalent-width measurements
of the Fe~\textsc{i} lines. Non-LTE (NLTE) effects might be present in the
Fe~\textsc{i} lines, which can affect the derived metallicity
\citep{lind2012}, but no Fe~\textsc{ii} lines were detected in any of the four
program stars. The measured Fe abundance may also be subject to uncertainties
from three-dimensional (3D) effects. \citet{collet2006} report a 3D correction
of $\sim -0.2$ dex for the Fe abundance for two of the most metal-poor stars
known (HE~0107-5240 and HE~1327$-$2326), both of which have temperatures and
gravities that are comparable, within the combined error bars, to those of the
stars presented in this paper. A better basis for comparison, at the
same metallicity as our program stars, is clearly desirable.
\citet{bergemann2012} found, however, that departures from LTE will
likely partly compensate such 3D LTE effects, leaving a smaller net
effect. Our stars have several Fe~\textsc{i} lines in common with the study
of \citet{bergemann2012}. A full 3D NLTE study is clearly warranted,
but beyond the scope of the present study.

The microturbulent velocity was computed in the usual way, by forcing
the abundances from Fe~\textsc{i} lines to show no trend with reduced equivalent
width, $\log(W_{\lambda}/\lambda)$. For HE~0233$-$0343, too few Fe~\textsc{i}
lines were present to determine the microturbulent velocity in this way,
so a fixed valued of $\xi = 2$~km s$^{-1}$ was used for this star.

For the warmer stars, HE~0233$-$0343 and HE~2239$-$5019, two possible
solutions for the surface gravity were found. Several tests were made to
settle on the listed values, both consistent with subgiant, rather than
dwarf, classifications. This aspect will be explored further in Paper~II
of this series. The final stellar parameters and their associated
uncertainties are listed in Table~\ref{tab1}.

\subsection{Abundance Analysis}

The abundance analysis has been carried out by synthesizing individual
spectral lines with the 2011 version of MOOG \citep{sneden1973}, which
includes a proper treatment of continuum scattering \citep{sobeck2011}.
A set of $\alpha$-enhanced ATLAS9 models \citep{castelli2003} have been
used, along with interpolation software tested in \citet{allende2004},
which produces models with the required stellar parameters \citep[e.g.,
][]{reddy2003,allende2004}. For HE~0233$-$0343, the metallicity in the
model atmosphere was $\mathrm{[m/H]} = -4.5$, which differs by 0.18 dex
from the metallicity of the star. This difference is within the
uncertainty of the derived $\mathrm{[Fe/H]}$ of the star and given the
small difference, we expect no change in any of the abundances when
using a model with $\mathrm{[m/H]} = -4.7$.
 
The {\it Gaia}/ESO line list version 3 has been used (Heiter et al., in
preparation). Atomic data from VALD \citep{kupka2000} were adopted for lines
not included in that line list. Hyperfine splitting was taken into account
for lines of Sc, Mn, and Co, using the data from \citet{kurucz1995}. For
Ba and Li, both hyperfine splitting and isotope shifts are present, and
data from \citet{mcwilliam1998} and \citet{asplund2006} were included,
respectively. The molecular information for CH, CN, and NH was kindly
provided from T. Masseron (private communication) 

The derived elemental abundances, along with propagated uncertainties
arising from the effects of uncertain stellar parameters, continuum
placement, and line information, are listed in Table~\ref{tab1}. The
adopted solar abundances are from \citet{asplund2009}. All listed
abundances are derived under one-dimentional (1D) and LTE assumptions. NLTE
effects will be explored in Paper~II.

\begin{deluxetable*}{lrrrr}
\tablecaption{Stellar Parameters and Derived Abundances \label{tab1}}
\tablewidth{0pt}
\tablehead{
\colhead{} & \colhead{\object{HE 0134$-$1519}} & \colhead{\object{HE 0233$-$0343}} &
\colhead{\object{HE 1310$-$0536}} & \colhead{\object{HE 2239$-$5019}}}
\startdata
R.A.  & 01 37 05.4 & 02 36 29.7 & 13 13 31.2 & 22 42 26.9\\
Decl. & $-$15 04 24  & $-03$ 30 06  & $-05$ 52 13  & $-50$ 04 01\\
$V$\tablenotemark{a}   & 14.47& 15.43& 14.35& 15.85\\
$B-V$\tablenotemark{a} & 0.50 & 0.34 & 0.71 & 0.39 \\
$J-K$\tablenotemark{a} & 0.43 & 0.30 & 0.64 & 0.40 \\
Radial velocity (km s$^{-1}$) & 244 & 64 & 113 & 370 \\
\cutinhead{Parameters}
$T_{\rm eff}$ ($\pm$100~K)     & 5500  & 6100  & 5000  & 6100 \\
$\log g$ ($\pm$0.3~dex)       & 3.2  & 3.4  & 1.9  & 3.5 \\
$[$Fe/H$]$ ($\pm$0.2~dex)     & $-4.0$ & $-4.7$ & $-4.2$ & $-4.2$ \\ 
$\xi$ ($\pm$0.3~km s$^{-1}$) & 1.5  & 2.0  &  2.2 & 1.8 \\
\cutinhead{Abundances}
$A$(Li)              &      1.27 (0.19)  &       1.77 (0.18)  &    $<$0.80\nodata  &$<$1.70\nodata \\
$\mathrm{[Fe/H]}$  &   $-$3.98 (0.30)  &    $-$4.68 (0.30)  &    $-$4.15 (0.30)  &   $-$4.15 (0.30)\\
$\mathrm{[C/Fe]}$  &   $+$1.00 (0.26)  &    $+$3.48 (0.24)  &    $+$2.36 (0.23)  &$<$$+$1.70\nodata \\
$\mathrm{[N/Fe]}$  &$<$$+$1.00\nodata  & $<$$+$2.80\nodata  &    $+$3.20 (0.37)  &$<$$+$2.70\nodata \\
$\mathrm{[Na/Fe]}$ &   $-$0.24 (0.15)  & $<$$+$0.50\nodata  &    $+$0.19 (0.14)  &$<$$-$0.30\nodata \\
$\mathrm{[Mg/Fe]}$ &   $+$0.25 (0.14)  &    $+$0.59 (0.15)  &    $+$0.42 (0.16)  & $+$0.45 (0.15) \\
$\mathrm{[Al/Fe]}$ &   $-$0.38 (0.20)  & $<$$+$0.03\nodata  &    $-$0.39 (0.21)  & $-$0.57 (0.21) \\
$\mathrm{[Si/Fe]}$ &   $+$0.05 (0.16)  &    $+$0.37 (0.15)  & $<$$+$0.25\nodata  & $+$0.06 (0.15) \\
$\mathrm{[Ca/Fe]}$ &   $+$0.10 (0.13)  &    $+$0.34 (0.15)  &       0.00 (0.20)  & $+$0.23 (0.15) \\
$\mathrm{[Sc/Fe]}$ &   $-$0.10 (0.18)  & $<$$+$0.20\nodata  &    $-$0.23 (0.16)  & $+$0.26 (0.16) \\
$\mathrm{[Ti/Fe]}$ &   $+$0.11 (0.21)  &    $+$0.18 (0.17)  &    $+$0.35 (0.18)  & $+$0.37 (0.17) \\
$\mathrm{[Cr/Fe]}$ &   $-$0.22 (0.18)  & $<$$+$0.50\nodata  &    $-$0.49 (0.26)  &    0.00 (0.17) \\
$\mathrm{[Mn/Fe]}$ &   $-$1.19 (0.19)  & $<$$-$0.10\nodata  &    $-$1.40 (0.20)  &$<$$-$0.60\nodata \\
$\mathrm{[Co/Fe]}$ &   $+$0.25 (0.18)  & $<$$+$1.60\nodata  &    $+$0.10 (0.16)  &$<$$+$0.70\nodata \\
$\mathrm{[Ni/Fe]}$ &   $+$0.19 (0.19)  & $<$$+$0.90\nodata  &    $-$0.12 (0.20)  & $+$0.24 (0.17) \\
$\mathrm{[Sr/Fe]}$ &   $-$0.30 (0.19)  &     $+$0.32 (0.19) &    $-$1.08 (0.14)  &$<$$-$0.60\nodata \\
$\mathrm{[Ba/Fe]}$ &$<$$-$0.50\nodata  & $<$$+$0.80\nodata  &    $-$0.50 (0.15)  & $<$0.00\nodata \\
\enddata
\tablenotetext{a}{\citet{beers2007}}
\end{deluxetable*}

\section{Results}

\subsection{Radial Velocity}

Two of the stars listed in Table~\ref{tab1}, HE~0134$-$1519 and
HE~2239$-$5019, exhibit quite high radial velocities, 244~km~s$^{-1}$
and 370~km~s$^{-1}$, respectively. The uncertainty of the listed radial
velocities is on the order of $\sim 1$~km~s$^{-1}$. Such high velocities
may suggest membership in the proposed outer-halo population of the
Milky Way \citep{carollo2007, carollo2010, beers2012}. A kinematic
analysis of the full space motions of our complete program sample,
including the four stars reported on here, will be presented in Paper~II
of this series. In this context, it is interesting that
\citet{carollo2014} present tentative evidence that the CEMP-$s$ and
CEMP-no stars may well be associated with progenitors that belong, in
different proportion, to the suggested inner- and outer-halo populations
of the Milky Way. 

\subsection{Elemental Abundances}

Our analysis has produced abundance estimates, or upper limits,
for 17 elements -- Li, C, N, Na, Mg, Al, Si, Ca, Sc, Ti, Cr, Mn, Fe, Co,
Ni, Sr, and Ba. We describe these analyses in detail in the subsections
below.

\subsubsection{Lithium}

\begin{figure}
\begin{center}
\includegraphics[angle=0,width=3.5in]{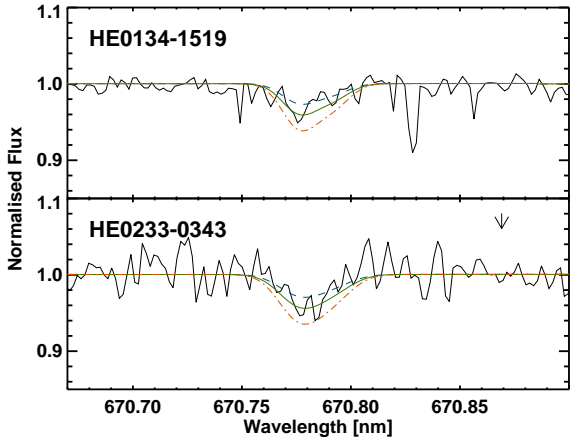}
\end{center}
\caption{Li line fit for HE~0134$-$1519 (top) $A$(Li) = 1.46,
1.27, and 1.08 (blue dashed line, solid green line, and red dot-dashed
line, respectively) and HE~0233$-$0343 (bottom) $A$(Li) =
1.95, 1.77, and 1.59 (blue dashed line, solid green line, and red
dot-dashed line, respectively). The blue dashed and red dot-dashed lines
  correspond to $A$(Li)$\pm\sigma$(Li), respectively, as listed in
  Table~\ref{tab1}. \label{fig2}} 
\end{figure}

We derived lithium abundances from synthesis of the Li~\textsc{i} 6707.8\,{\AA}
doublet. Lithium is detected for two of our program stars--HE~0134$-$1519, with
$A$(Li) = 1.27\footnote {$A$(Li) is defined in the usual manner, $A$(Li) $=
  \log(N(\mbox{Li})/N(\mbox{H})) + 12$.}, and HE~0233$-$0343, with $A$(Li) =
1.77. Figure \ref{fig2} shows the spectral region around the Li line for two
of our stars (top: HE~0134$-$1519, and bottom: HE~0233$-$0343), together with
three synthetic spectra computed with $A$(Li) = 1.46, 1.27, and 1.08,
respectively, for HE~0134$-$1519, and $A$(Li) = 1.95, 1.77, and 1.59,
respectively, for HE~0233$-$0343. HE~0233$-$0343 is the second most metal-poor
star with a detected lithium line, as lithium was also detected in the most
metal-poor star known, SMSS~J031300.36-670839.3 with $\mathrm{[Fe/H]} < -7$,
recently discovered by \citet{keller2014} ($A$(Li) = 0.7). Li is not detected
for the two remaining program stars; we computed upper limits of $A$(Li) $ <
0.8$ and $A$(Li) $ < 1.70$ for HE~1310$-$0536 and HE~2239$-$5019,
respectively. The very low upper limit detected in HE~1310$-$0536 is expected,
as this star is sufficiently evolved that it has undergone first dredge
up. Its convective zone likely extends down to layers in the atmosphere where
lithium has been destroyed by nuclear burning.

\begin{figure}
\begin{center}
\includegraphics[angle=0,width=3.5in]{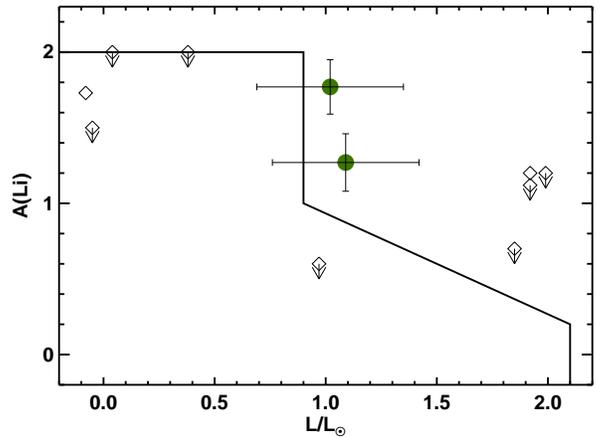}
\end{center}
\caption{LTE lithium abundances, $A$(Li), as a function of luminosity, for
  HE~0134$-$1519 and HE~0233$-$0343 (green circles), along with the CEMP-no
stars of \citet{masseron2012} (black diamonds). Upper limits are indicated by
arrows. The solid line indicates the division between Li-normal (above) and
Li-depleted (below) stars. \label{fig3}} 
\end{figure}

Figure \ref{fig3} displays the Li abundance for our two
CEMP-no stars with Li detections, as a function of their luminosity,
following Figure 16 of \citet{masseron2012}. Luminosities have been determined
in the same way as in \citet{masseron2012}, assuming $M=0.8M_\odot$. For
comparison, we also plot the CEMP-no stars of their sample. The solid line
marks the division between Li-normal (above) and Li-depleted (below)
stars. The line is computed from the Li abundance of non-CEMP stars with
luminosities in the range $-0.2<\log(L/L_\odot)<2.1$. The line follows the
Spite Li plateau for dwarf stars, then exhibits a linear decline in the Li
abundances of giants, where the Li is expected to be gradually depleted
due to convective burning episodes \citep[see ][ for
  details]{masseron2012}. Stars outside the above range in luminosity are
expected to have destroyed all their Li. Note that HE~1310$-$0536, with
$\log(L/L_\odot) = 2.11$, falls outside that range. Our two Li detections
both lie above the Li-normal line, but with lithium abundances below the Spite
plateau. Hence, Li has been depleted in these stars, consistent with the
result found by \citet{masseron2012}, that the CEMP-no class {\it only}
contains Li-depleted stars, even at these low metallicities. 

\subsubsection{Carbon}\label{carbon}

Three of our four program stars, HE~0134$-$1519, HE~0233$-$0343, and
HE~1310$-$0536, are carbon enhanced, with $\mathrm{[C/Fe]}\geq +0.7$.
They exhibit no enhancements in their neutron-capture elements
\citep[$\mathrm{[Ba/Fe]}\leq 0.0$;][]{beerschristlieb2005}, and are considered
CEMP-no stars. Technically, the status of HE~0233$-$0343 cannot be confirmed,
as only an upper limit for the Ba abundance of $\mathrm{[Ba/Fe]} < +0.8$ is
found. Considering that the great majority of CEMP stars with
$\mathrm{[Fe/H]}<-3$ are CEMP-no stars \citep{aoki2010}, and the fact that
there are no known CEMP-$s$ stars with [Fe/H] $< -3.5$, there is a high
likelihood that HE~0233$-$0343 also belongs to the CEMP-no class. The last of
the four stars, HE~2239$-$5019, shows no clear carbon enhancement; we compute
an upper limit of $\mathrm{[C/Fe]} < +1.7$ for this star. With no carbon
detected, this star is a potential candidate to be in the same class as
SDSS~J102915+172927, the only star with $\mathrm{[Fe/H]} < -4.5$ found not to
be carbon enhanced \citep{caffau2011}. 

\begin{figure}
\begin{center}
\includegraphics[angle=0,width=3.5in]{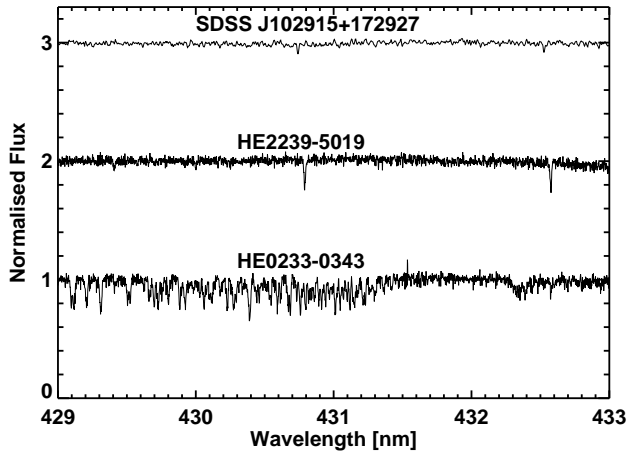}
\end{center}
\caption{Spectral range including the CH $G$ band in the spectra of
SDSS~J102915+172927 (top), HE~2239$-$5019 (middle), and HE~0233$-$0343
(bottom). The carbon lines are clearly seen in the spectrum of
HE~0233$-$0343, but are absent in the other two spectra.\label{fig4}}
\end{figure}

Figure~\ref{fig4} shows the spectral range including the CH
$G$ band for SDSS~J102915+172927, HE~2239$-$5019, and HE~0233$-$0343.
HE~0233$-$0343 has similar stellar parameters as HE~2239$-$5019, but it is
more iron poor and carbon enhanced. Similar to SDSS~J102915+172927, no
CH features are visible in HE~2239$-$5019. However, the noise level in the
spectrum of HE~2239$-$5019 is quite high, resulting in a high derived
upper limit on the carbon abundance, so it cannot be ruled out as being
a CEMP star.

Since three out of the four stars are carbon enhanced, the oxygen and
nitrogen abundances are also of interest. Nitrogen was detected in only
one star, HE~1310$-$0536, where the abundance listed in Table~\ref{tab1}
is derived from synthesis of the CN band at $3883$\,{\AA}. For the
remaining three stars, upper limits are derived from synthesis of the NH
band at $3360$\,{\AA}. Previous studies, such as \citet{sivarani2006}
and \citet{norris2013b}, have found a correlation of $\mathrm{[N/Fe]}$
with $\mathrm{[C/Fe]}$ for CEMP stars. The N abundance and upper limits
that we derive support this correlation. Oxygen was not detected in any of our
program stars, and the noise levels in the spectra were too high to compute a
meaningful upper limit on its abundance.

\subsubsection{Light Elements and Neutron-Capture Elements}

\begin{figure}
\begin{center}
\includegraphics[angle=0,width=3.5in]{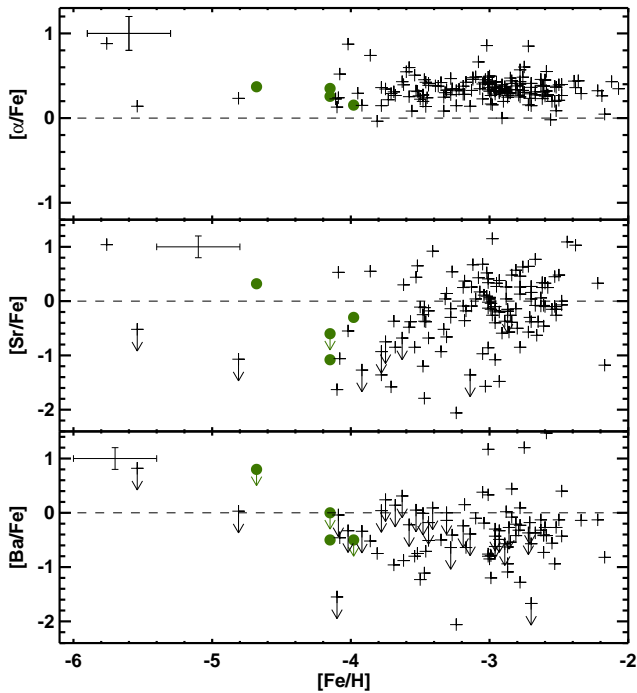}
\end{center}
\caption{Mean [$\alpha$/Fe] (top), [Sr/Fe] (middle), and [Ba/Fe] (bottom)
abundances for our four UMP stars (green circles) and the sample of
\citet{yong2013} (black crosses). Upper limits are indicated by arrows; the
dashed line is the solar value.\label{fig5}} 
\end{figure}

Since the stars in this sample have been analyzed in a similar manner as
those of \citet{yong2013}, the two samples are directly comparable. In
the top panel of Figure \ref{fig5}, the mean [$\alpha$/Fe] (taken to be
the mean of $\mathrm{[Mg/Fe]}$, $\mathrm{[Ca/Fe]}$, and
$\mathrm{[Ti/Fe]}$) abundance ratios of our four stars is compared to
those of \citet{yong2013}. Their sample includes some of the most
metal-poor stars known to date (HE~0107$-$5240: \citet{christlieb2002};
HE~1327$-$2326: \citet{frebel2005}; and HE~0557$-$4840: \citet{norris2007}.
A small over-abundance of the [$\alpha$/Fe] ratio is seen in the four
new stars, consistent with the existing picture of the $\alpha$-element
abundances in metal-poor stars, reflecting the enrichment from
core--collapse SNe in the early universe. \citet{norris2013b}
found that 50\% of their CEMP stars are more enhanced in the light
elements Na, Mg, Al, and Si, compared to other (C-normal) EMP stars with
similar stellar parameters. Among our program stars, HE~0233$-$0343
exhibits higher abundances of these elements relative to the rest of the
sample. However, none of our stars show over-abundances of
these elements as large as those found for some CEMP stars in the sample of
\citet{norris2013b}. The observed abundances for Al and Mn in our four
stars lie somewhat below the level predicted by the Galactic chemical
evolution models of \citet{nomoto2013}. This may be due to NLTE effects.
\citet{gehren2004} report NLTE corrections of +0.5~dex for Al in a
sample of metal-poor turn-off stars, while \citet{bergemann2008} find
corrections of up to +0.7~dex for Mn in their sample of metal-poor giant
and dwarf stars. This would bring Al to the predicted level, whereas Mn
would stay just below. 

The middle and bottom panels of Figure \ref{fig5} display the
$\mathrm{[Sr/Fe]}$ and $\mathrm{[Ba/Fe]}$ abundance ratios,
respectively, as functions of metallicity for our program stars and
those of \citet{yong2013}. Both samples exhibit a large spread in the
$\mathrm{[Sr/Fe]}$ and $\mathrm{[Ba/Fe]}$ ratios. The spread of
abundances for these two elements was also discussed by
\citet{chansen2012,chansen2013} and \citet{ yong2013}, all suggesting that
more than one production site exists for Sr and Ba. The scatter in the Sr and
Ba abundances of EMP stars has also been discussed by \citet{aoki2013}, who
studied the [Sr/Ba] ratios in a sample of 260 EMP stars. They detected no
stars with $\mathrm{[Sr/Fe]} > 0.0$ for $\mathrm{[Fe/H]} < -3.6$ (note
that their sample only includes four stars with $\mathrm{[Fe/H]} <
-3.6$). They proposed to explain the distribution in the observed
[Sr/Ba] ratios with a truncated $r$-process taking place in a type II SN, as
described by \citet{boyd2012}. \citet{aoki2013} also stated that neither the
$r$ process nor the truncated $r$ process are expected to produce stars with
$\mathrm{[Sr/Ba]} < -0.5$. They find six stars in their sample with
$\mathrm{[Sr/Ba]} < -0.5$, but suspect these to be contaminated with
$s$-process material. 

\begin{figure}
\begin{center}
\includegraphics[angle=0,width=3.5in]{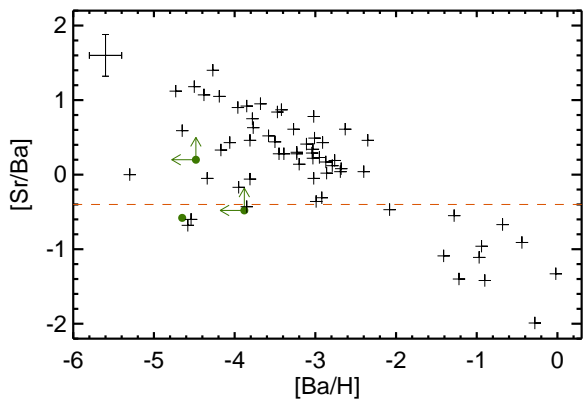}
\end{center}
\caption{$\mathrm{[Sr/Ba]}$ ratios plotted against Ba abundances,
  $\mathrm{[Ba/H]}$, for our three CEMP-no stars (green circles) and the sample
  of \citet{yong2013} (black crosses). Arrows indicate upper limits; the dashed
  red line indicates $\mathrm{[Sr/Ba]=-0.4}$. Ratios above this line indicate
  production of Sr and Ba by the weak $s$ process in massive stars or by the
  $r$ process, while those below indicate production by the main $s$ process
  in AGB stars.\label{fig6}} 
\end{figure}

\section{Discussion}

The lithium abundances in carbon-enhanced stars is a relatively
unexplored chapter in the history of Galactic chemical evolution;
theoretical efforts include \citet{stancliffe2009}. Only a few CEMP
stars have detected lithium and even fewer of these are CEMP-no stars,
though the samples of CEMP-no stars are increasing quickly, in
particular, from dedicated searches for CEMP stars \citep[e.g.,
][]{placco2010,placco2011,placco2013,placco2014}. We have detected Li
for two of the stars in our sample and the derived Li abundances for
these indicate a Li depletion in the stars relative to the Spite Li
plateau. These detections highlight the need for a progenitor of CEMP-no
stars that produces large amounts of carbon, but only small amounts of
neutron-capture elements, while to some extent depleting the lithium.
\citet{masseron2012} test how mass transfer from an AGB companion will
affect the Li abundance of a CEMP star. They examined a set of different
AGB models and different depletion factors for the transferred material,
but found that none of the models could explain the observed spread in
Li abundances of the CEMP-no stars of their sample. The other suggested
progenitor candidates for the CEMP-no stars include faint SNe
that experienced mixing and fallback, as well as spinstars. If these are
indeed the progenitors, the Li abundance of CEMP-no stars should lie
below the level found in non-carbon-enhanced stars, as Li should be
depleted (or totally destroyed) in such objects. Hence, when the gas
from these mixes with the ISM in their surroundings (and forms the
CEMP-no stars), the overall Li abundance will be lowered
\citep{meynet2010}. In fact, as suggested by \citet{piau2006}, this
process might be responsible for the lowering of the primordial Li
abundance from the level predicted from big bang nucleosynthesis
calculations, the lack of scatter among stars on the plateau at
metallicities $-2.5 <  \mathrm{[Fe/H]} < -1.5$, due to complete mixing
\citep[e.g., ][]{ryan1999}, the downturn and increase of scatter in the
Li abundances for stars with$ \mathrm{[Fe/H]} < -2.5$, due to incomplete local
mixing \citep{sbordone2010}, and the very low (or absent) Li among the
lowest metallicity stars \citep[e.g., ][]{frebel2005,keller2014}.
 
The sample of Li measurements for CEMP-no stars is presently very
limited, and at this stage all of the proposed progenitors of CEMP-no
stars involve some variation of mixing. When mixing has occurred, it is
natural that the Li abundance is depleted, leading to lower abundances
of lithium in carbon-enhanced stars. Also, it is uncertain how much of
the Li can be depleted after a possible mass transfer via mixing and
rotation of the CEMP star itself \citep{talon2005,stancliffe2007}. More
Li detections (or strong upper limits) in CEMP-no stars are needed in
order to better understand the nature of the progenitors of these stars.

The carbon enhancement, detected for three of our four program stars 
is consistent with the picture of carbon enhancement in the early
universe found by other authors \citep[e.g.,][]{carollo2012,
norris2013b}. An enrichment of carbon in the early universe also
supports one of the proposed formation scenarios for low-mass stars,
that gas clouds can fragment as a result of cooling via fine-structure
lines of carbon and oxygen \citep{frebel2007}. 

\citet{spite2013} examined the carbon abundances of
dwarfs and turnoff stars, stars in which mixing has not altered the
carbon abundance at the surface of the star. From their sample, they
suggested the presence of two plateaus of the carbon abundances, one for
$\mathrm{[Fe/H]} > -3.0$ at $A$(C) $\sim$ 8.25 and one for
$\mathrm{[Fe/H]} < -3.4$ at $A$(C) $\sim$ 6.8. They point to the
low number of stars observed with $\mathrm{[Fe/H]} < -3.4$, and highlight the
difficulty of observing carbon in warmer, unmixed stars. As a result, they
could not conclude if the lower plateau is just an upper limit on the
detections or an actual plateau. We derive a carbon abundance of
$A$(C) = 7.23 for the unmixed CEMP subgiant in our sample
HE~0233$-$0343, placing it a little above the plateau. Clearly, observations
of additional unmixed CEMP stars are needed to clarify if such a plateau
exists. 

The origin of neutron-capture elements in low-metallicity stars is not
yet well-understood. A large spread is seen in the abundances of the
neutron-capture elements Sr and Ba for CEMP-no stars (indistinguishable
from that of non-carbon-rich metal-poor stars). For the CEMP-$s$ stars,
the carbon and $s$-process overabundances are believed to be the result
of mass transfer from a binary AGB companion. Indeed,
\citet{lucatello2005} showed that a significant fraction of these stars
(perhaps all) are in fact in binary systems. For the CEMP-no stars,
however, early results from radial-velocity monitoring do not require
them to be in binary systems
(\citet{cohen2013,thansen2013,norris2013b,starkenburg2014}; J. Andersen et al.,
in prep.).

Figure \ref{fig6} shows the derived $\mathrm{[Sr/Ba]}$ ratios of our
three CEMP-no stars, together with the ratios for stars from the
\citet{yong2013} sample that had detections of both Sr and Ba. The
dashed red line indicates $\mathrm{[Sr/Ba]=-0.4}$, used as an upper
limit for the main $s$-process signature of AGB stars
\citep{spite2013}. At solar metallicity, Sr is a tracer of the weak
$s$ process, in massive stars \citep{heil2009,pignatari2010}, while Ba is
a tracer of the main $s$ process taking place in AGB stars
\citep{busso1999, kappeler2011}. At low metallicity, where the main
$s$ process is not yet active, the picture is different.

To assess the origin of the Sr and Ba detected for our three CEMP-no
stars, the $\mathrm{[Sr/Ba]}$ ratio can be compared to that for
classical main $s$-process-enhanced metal-poor stars, and in
strongly $r$-process-enhanced metal-poor stars.
\citet{lucatello2003} reported on the abundances analysis of
HE~0024$-$2523, a classical main $s$-process-enhanced star with
carbon enhancement. This star was also found to be in a binary system,
and the authors argued that the carbon and $s$-process-element
enhancement is the result of mass transfer from an AGB companion. The
$\mathrm{[Sr/Ba]}$ ratio in this star is $\mathrm{[Sr/Ba]}=-1.12$, a
very low value, due to its high Ba abundance. Although a large spread is
seen in the efficiency of the main $s$-process element production
of the AGB stars \citep{bisterzo2011}, a low $\mathrm{[Sr/Ba]}$ ratio is
observed for $s$-process elements produced in AGB stars;
\citet{spite2013} use $\mathrm{[Sr/Ba] < -0.4}$ as an upper limit, while
$\mathrm{[Sr/Ba]=-0.5}$ was used by \citet{aoki2013}. For our CEMP-no
stars, we find the following [Sr/Ba] ratios: $\mathrm{[Sr/Ba]} > +0.20$
(HE~0134$-$1519), $\mathrm{[Sr/Ba]} > -0.48$ (HE~0233$-$0343), and
$\mathrm{[Sr/Ba]} = -0.58$ (HE~1310$-$0536). The ratios found in
HE~0233$-$0343 and HE~1310$-$0536 could indicate production by the main
$s$ process. However, these stars are CEMP-no stars, i.e., their individual
abundance ratios of Ba relative to iron are low ($\mathrm{[Ba/Fe]} <
0$), and they are also UMP stars ($\mathrm{[Fe/H]} < -4.0$). At such low
metallicity, the Ba is more likely produced in the main $r$ process from
SNe and Sr in the weak $s$ process in massive stars. The following
$\mathrm{[Sr/Ba]}$ ratios have been found in strongly $r$-process-enhanced
metal-poor stars, $\mathrm{[Sr/Ba]}=-0.52$ in CS~31082-001 \citep{hill2002},
$\mathrm{[Sr/Ba]}=-0.41$ for CS~22892$-$052 \citep{sneden2003}, and
$\mathrm{[Sr/Ba]}=-0.46$ for CS~29497$-$004 \citep{christlieb2004}. These ratios
are, very similar to those we find, for HE~0233$-$0343 and HE~1310$-$0536. The
$\mathrm{[Sr/Ba]}$ ratio found for HE~0134$-$1519 indicates that the Sr and Ba
in this star could have been produced in the weak $s$ process in spinstars. 

\citet{cescutti2013} proposed that the
spread in Sr and Ba abundances detected in CEMP-no stars could be
explained by spinstar progenitors. Their model includes a standard
$r$ process (presumably in the natal clouds) plus a contribution from
the weak $s$ process occurring in spinstars. With this combination,
they can model the spread seen in the abundances of Sr and Ba in
metal-poor stars, including the CEMP-no stars, while also reproducing
the low scatter in $\alpha$-elements. They do, however, state that their
models cannot reproduce the $\mathrm{[C/O]}$ and $\mathrm{[N/O]}$ ratios
in the same CEMP-no stars, but point to the scenario of
\citet{meynet2010}, where low-mass stars belonging to the forming
stellar cluster of a spinstar are enriched in carbon via stellar winds
from the spinstar.

\section{Summary}

We have conducted a detailed chemical-abundance analysis of
four new UMP stars, with $\mathrm{[Fe/H]} \leq -4.0$; fewer than 10
such stars were previously known. The Li, C, Sr, and Ba measurements
provide a new observational window to examine nucleosynthesis at the
earliest times in our Galaxy. While one star has an upper limit of
$\mathrm{[C/Fe]} < +1.7$, the remaining three stars are all C-rich, confirming
the prevalence of CEMP stars in the early universe. The detection of
Li in two of the clear CEMP-no stars requires that whatever process(es)
produce(s) the large amount of C (and presumably the N, O often found in
CEMP-no stars, but which could not be detected in our stars), does not
completely destroy Li. In light of the newer data for C and Li for
these, and other recently studied CEMP-no stars, we suggest that it is
worth revisiting the Li astration model described by \citet{piau2006}.
Finally, our detections of Sr and Ba for several additional UMP stars
demonstrates that the process(es) creating these elements are at work even at
very low metallicities, a conclusion also reached by
\citet{roederer2013}. Since there still remain a number of stars at the lowest
metallicities with only upper limits on Sr and/or Ba, increasing the sample
sizes and the quality of the available high-resolution spectroscopy for stars
at these metallicities is an essential step toward understanding
nucleosynthesis at the earliest epochs and ultimately to characterize ``the
frequency and environmental influence of the astrophysical sites of
heavy-element production'' \citep{roederer2013}.

\acknowledgments

We thank E. Caffau for providing the spectrum of SDSS~J102915+172927.
This work was supported by Sonderforschungsbereich SFB 881 ``The Milky
Way System'' (subproject A4) of the German Research Foundation (DFG).
T.C.B. acknowledges partial support from grant PHY 08-22648: Physics
Frontier Center/Joint Institute for Nuclear Astrophysics (JINA),
awarded by the U.S. National Science Foundation. V.M.P. acknowledges
support from the Gemini Observatory. M.A., M.S.B., J.E.N., and D.Y.
acknowledge support from the Australian Research Council (grants
DP0342613, DP0663562, and FL110100012) for studies of the Galaxy's most
metal-poor stars. A.F. acknowledges support from NSF grant AST-1255160.
Furthermore, we thank the referee for helpful comments.

This work made use of the Southern Astrophysical Research (SOAR)
4.1 m Telescope (proposal SO2013B-001), 
which is a joint project of the Minist\'{e}rio da Ci\^{e}ncia, Tecnologia, e 
Inova\c{c}\~{a}o (MCTI) da Rep\'{u}blica Federativa do Brasil, the U.S. 
National Optical Astronomy Observatory (NOAO), the University of North
Carolina at  Chapel Hill (UNC), and Michigan State University (MSU).

This work also made use of the Kitt Peak National Observatory Mayall 4 m
telescope (proposal 2013B-0046) of the National Optical Astronomy Observatory, 
which is operated by the Association of American Universities for Research in
Astronomy (AURA), under cooperative agreement with the National Science
Foundation. 

This work also made use of observations conducted with ESO
Telescopes at the La Silla Observatory and the VLT on Paranal, under
programme ID 69.D-0130(A).

\bibliographystyle{apj}
\bibliography{THansen}

\begin{thebibliography}{}
\expandafter\ifx\csname natexlab\endcsname\relax\def\natexlab#1{#1}\fi

\bibitem[{{Allende Prieto} {et~al.}(2004){Allende Prieto}, {Barklem},
  {Lambert}, \& {Cunha}}]{allende2004}
{Allende Prieto}, C., {Barklem}, P.~S., {Lambert}, D.~L., \& {Cunha}, K. 2004,
  Astronomy and Astrophysics, 420, 183

\bibitem[{Aoki(2009)}]{aoki2010}
Aoki, W. 2009, Proceedings of the International Astronomical Union, 5, 111

\bibitem[{{Aoki} {et~al.}(2007){Aoki}, {Beers}, {Christlieb}, {Norris}, {Ryan},
  \& {Tsangarides}}]{aoki2007}
{Aoki}, W., {Beers}, T.~C., {Christlieb}, N., {et~al.} 2007, \apj, 655, 492

\bibitem[{{Aoki} {et~al.}(2013){Aoki}, {Suda}, {Boyd}, {Kajino}, \&
  {Famiano}}]{aoki2013}
{Aoki}, W., {Suda}, T., {Boyd}, R.~N., {Kajino}, T., \& {Famiano}, M.~A. 2013,
  The Astrophysical Journal Letters, 766, L13

\bibitem[{{Asplund} {et~al.}(2009){Asplund}, {Grevesse}, {Sauval}, \&
  {Scott}}]{asplund2009}
{Asplund}, M., {Grevesse}, N., {Sauval}, A.~J., \& {Scott}, P. 2009, Annual
  Review of Astronomy \& Astrophysics, 47, 481

\bibitem[{{Asplund} {et~al.}(2006){Asplund}, {Lambert}, {Nissen}, {Primas}, \&
  {Smith}}]{asplund2006}
{Asplund}, M., {Lambert}, D.~L., {Nissen}, P.~E., {Primas}, F., \& {Smith},
  V.~V. 2006, The Astrophysical Journal, 644, 229

\bibitem[{{Beers} \& {Christlieb}(2005)}]{beerschristlieb2005}
{Beers}, T.~C., \& {Christlieb}, N. 2005, Annual Review of Astronomy \&
  Astrophysics, 43, 531

\bibitem[{{Beers} {et~al.}(2007){Beers}, {Flynn}, {Rossi}, {Sommer-Larsen},
  {Wilhelm}, {Marsteller}, {Lee}, {De Lee}, {Krugler}, {Deliyannis}, {Simmons},
  {Mills}, {Zickgraf}, {Holmberg}, {{\"O}nehag}, {Eriksson}, {Terndrup},
  {Salim}, {Andersen}, {Nordstr{\"o}m}, {Christlieb}, {Frebel}, \&
  {Rhee}}]{beers2007}
{Beers}, T.~C., {Flynn}, C., {Rossi}, S., {et~al.} 2007, \apjs, 168, 128

\bibitem[{{Beers} {et~al.}(2012){Beers}, {Carollo}, {Ivezi{\'c}}, {An},
  {Chiba}, {Norris}, {Freeman}, {Lee}, {Munn}, {Re Fiorentin}, {Sivarani},
  {Wilhelm}, {Yanny}, \& {York}}]{beers2012}
{Beers}, T.~C., {Carollo}, D., {Ivezi{\'c}}, {\v Z}., {et~al.} 2012, \apj, 746,
  34

\bibitem[{{Bergemann} \& {Gehren}(2008)}]{bergemann2008}
{Bergemann}, M., \& {Gehren}, T. 2008, \aap, 492, 823

\bibitem[{{Bergemann} {et~al.}(2012){Bergemann}, {Lind}, {Collet}, {Magic}, \&
  {Asplund}}]{bergemann2012}
{Bergemann}, M., {Lind}, K., {Collet}, R., {Magic}, Z., \& {Asplund}, M. 2012,
  \mnras, 427, 27

\bibitem[{{Bessell}(2007)}]{bessell2007}
{Bessell}, M.~S. 2007, \pasp, 119, 605

\bibitem[{{Bisterzo} {et~al.}(2011){Bisterzo}, {Gallino}, {Straniero},
  {Cristallo}, \& {K{\"a}ppeler}}]{bisterzo2011}
{Bisterzo}, S., {Gallino}, R., {Straniero}, O., {Cristallo}, S., \&
  {K{\"a}ppeler}, F. 2011, \mnras, 418, 284

\bibitem[{{Boyd} {et~al.}(2012){Boyd}, {Famiano}, {Meyer}, {Motizuki},
  {Kajino}, \& {Roederer}}]{boyd2012}
{Boyd}, R.~N., {Famiano}, M.~A., {Meyer}, B.~S., {et~al.} 2012, The
  Astrophysical Journal Letters, 744, L14

\bibitem[{{Burbidge} {et~al.}(1957){Burbidge}, {Burbidge}, {Fowler}, \&
  {Hoyle}}]{burbidge1957}
{Burbidge}, E.~M., {Burbidge}, G.~R., {Fowler}, W.~A., \& {Hoyle}, F. 1957,
  Reviews of Modern Physics, 29, 547

\bibitem[{{Busso} {et~al.}(1999){Busso}, {Gallino}, \&
  {Wasserburg}}]{busso1999}
{Busso}, M., {Gallino}, R., \& {Wasserburg}, G.~J. 1999, \araa, 37, 239

\bibitem[{{Caffau} {et~al.}(2011){Caffau}, {Bonifacio}, {Fran{\c c}ois},
  {Sbordone}, {Monaco}, {Spite}, {Spite}, {Ludwig}, {Cayrel}, {Zaggia},
  {Hammer}, {Randich}, {Molaro}, \& {Hill}}]{caffau2011}
{Caffau}, E., {Bonifacio}, P., {Fran{\c c}ois}, P., {et~al.} 2011, Nature, 477,
  67

\bibitem[{{Carollo} {et~al.}(2014){Carollo}, {Freeman}, {Beers}, {Placco},
  {Tumlinson}, \& {Martell}}]{carollo2014}
{Carollo}, D., {Freeman}, K.~C., {Beers}, T.~C., {et~al.} 2014, The
  Astrophysical Journal, arXiv:1401.0574

\bibitem[{{Carollo} {et~al.}(2007){Carollo}, {Beers}, {Lee}, {Chiba}, {Norris},
  {Wilhelm}, {Sivarani}, {Marsteller}, {Munn}, {Bailer-Jones}, {Fiorentin}, \&
  {York}}]{carollo2007}
{Carollo}, D., {Beers}, T.~C., {Lee}, Y.~S., {et~al.} 2007, \nat, 450, 1020

\bibitem[{{Carollo} {et~al.}(2010){Carollo}, {Beers}, {Chiba}, {Norris},
  {Freeman}, {Lee}, {Ivezi{\'c}}, {Rockosi}, \& {Yanny}}]{carollo2010}
{Carollo}, D., {Beers}, T.~C., {Chiba}, M., {et~al.} 2010, \apj, 712, 692

\bibitem[{{Carollo} {et~al.}(2012){Carollo}, {Beers}, {Bovy}, {Sivarani},
  {Norris}, {Freeman}, {Aoki}, {Lee}, \& {Kennedy}}]{carollo2012}
{Carollo}, D., {Beers}, T.~C., {Bovy}, J., {et~al.} 2012, The Astrophysical
  Journal, 744, 195

\bibitem[{{Castelli} \& {Kurucz}(2003)}]{castelli2003}
{Castelli}, F., \& {Kurucz}, R.~L. 2003, in IAU Symposium, Vol. 210, Modelling
  of Stellar Atmospheres, ed. N.~{Piskunov}, W.~W. {Weiss}, \& D.~F. {Gray},
  20P

\bibitem[{{Cescutti} {et~al.}(2013){Cescutti}, {Chiappini}, {Hirschi},
  {Meynet}, \& {Frischknecht}}]{cescutti2013}
{Cescutti}, G., {Chiappini}, C., {Hirschi}, R., {Meynet}, G., \&
  {Frischknecht}, U. 2013, \aap, 553, A51

\bibitem[{{Christlieb} {et~al.}(2002){Christlieb}, {Bessell}, {Beers},
  {Gustafsson}, {Korn}, {Barklem}, {Karlsson}, {Mizuno-Wiedner}, \&
  {Rossi}}]{christlieb2002}
{Christlieb}, N., {Bessell}, M.~S., {Beers}, T.~C., {et~al.} 2002, Nature, 419,
  904

\bibitem[{{Christlieb} {et~al.}(2004){Christlieb}, {Beers}, {Barklem},
  {Bessell}, {Hill}, {Holmberg}, {Korn}, {Marsteller}, {Mashonkina}, {Qian},
  {Rossi}, {Wasserburg}, {Zickgraf}, {Kratz}, {Nordstr{\"o}m}, {Pfeiffer},
  {Rhee}, \& {Ryan}}]{christlieb2004}
{Christlieb}, N., {Beers}, T.~C., {Barklem}, P.~S., {et~al.} 2004, \aap, 428,
  1027

\bibitem[{{Cohen} {et~al.}(2013){Cohen}, {Christlieb}, {Thompson}, {McWilliam},
  {Shectman}, {Reimers}, {Wisotzki}, \& {Kirby}}]{cohen2013}
{Cohen}, J.~G., {Christlieb}, N., {Thompson}, I., {et~al.} 2013, \apj, 778, 56

\bibitem[{{Collet} {et~al.}(2006){Collet}, {Asplund}, \&
  {Trampedach}}]{collet2006}
{Collet}, R., {Asplund}, M., \& {Trampedach}, R. 2006, \apjl, 644, L121

\bibitem[{{Dekker} {et~al.}(2000){Dekker}, {D'Odorico}, {Kaufer}, {Delabre}, \&
  {Kotzlowski}}]{dekker2000}
{Dekker}, H., {D'Odorico}, S., {Kaufer}, A., {Delabre}, B., \& {Kotzlowski}, H.
  2000, in Society of Photo-Optical Instrumentation Engineers (SPIE) Conference
  Series, Vol. 4008, Optical and IR Telescope Instrumentation and Detectors,
  ed. M.~{Iye} \& A.~F. {Moorwood}, 534--545

\bibitem[{{Demarque} {et~al.}(2004){Demarque}, {Woo}, {Kim}, \&
  {Yi}}]{demarque2004}
{Demarque}, P., {Woo}, J.-H., {Kim}, Y.-C., \& {Yi}, S.~K. 2004, The
  Astrophysical Journals, 155, 667

\bibitem[{{Dopita} {et~al.}(2007){Dopita}, {Hart}, {McGregor}, {Oates},
  {Bloxham}, \& {Jones}}]{dopita2007}
{Dopita}, M., {Hart}, J., {McGregor}, P., {et~al.} 2007, \apss, 310, 255

\bibitem[{{Frebel} {et~al.}(2007){Frebel}, {Johnson}, \& {Bromm}}]{frebel2007}
{Frebel}, A., {Johnson}, J.~L., \& {Bromm}, V. 2007, \mnras, 380, L40

\bibitem[{{Frebel} {et~al.}(2005){Frebel}, {Aoki}, {Christlieb}, {Ando},
  {Asplund}, {Barklem}, {Beers}, {Eriksson}, {Fechner}, {Fujimoto}, {Honda},
  {Kajino}, {Minezaki}, {Nomoto}, {Norris}, {Ryan}, {Takada-Hidai},
  {Tsangarides}, \& {Yoshii}}]{frebel2005}
{Frebel}, A., {Aoki}, W., {Christlieb}, N., {et~al.} 2005, Nature, 434, 871

\bibitem[{{Gehren} {et~al.}(2004){Gehren}, {Liang}, {Shi}, {Zhang}, \&
  {Zhao}}]{gehren2004}
{Gehren}, T., {Liang}, Y.~C., {Shi}, J.~R., {Zhang}, H.~W., \& {Zhao}, G. 2004,
  \aap, 413, 1045

\bibitem[{{Gustafsson} {et~al.}(2008){Gustafsson}, {Edvardsson}, {Eriksson},
  {J{\o}rgensen}, {Nordlund}, \& {Plez}}]{gustafsson2008}
{Gustafsson}, B., {Edvardsson}, B., {Eriksson}, K., {et~al.} 2008, Astronomy
  and Astrophysics, 486, 951

\bibitem[{{Hansen} {et~al.}(2013{\natexlab{a}}){Hansen}, {Bergemann},
  {Cescutti}, {Fran{\c c}ois}, {Arcones}, {Karakas}, {Lind}, \&
  {Chiappini}}]{chansen2013}
{Hansen}, C.~J., {Bergemann}, M., {Cescutti}, G., {et~al.} 2013{\natexlab{a}},
  \aap, 551, A57

\bibitem[{{Hansen} {et~al.}(2012){Hansen}, {Bergemann}, {Cescutti}, {Francois},
  {Arcones}, {Karakas}, {Lind}, \& {Chiappini}}]{chansen2012}
---. 2012, ArXiv e-prints, arXiv:1212.4147

\bibitem[{{Hansen} {et~al.}(2011){Hansen}, {Andersen}, {Nordstr{\"o}m},
  {Buchhave}, \& {Beers}}]{thansen2011}
{Hansen}, T., {Andersen}, J., {Nordstr{\"o}m}, B., {Buchhave}, L.~A., \&
  {Beers}, T.~C. 2011, \apjl, 743, L1

\bibitem[{{Hansen} {et~al.}(2013{\natexlab{b}}){Hansen}, {Andersen}, \&
  {Nordtr{\"o}m}}]{thansen2013}
{Hansen}, T., {Andersen}, J., \& {Nordtr{\"o}m}, B. 2013{\natexlab{b}}, ArXiv
  e-prints, arXiv:1301.7208

\bibitem[{{Heil} {et~al.}(2009){Heil}, {Juseviciute}, {K{\"a}ppeler},
  {Gallino}, {Pignatari}, \& {Uberseder}}]{heil2009}
{Heil}, M., {Juseviciute}, A., {K{\"a}ppeler}, F., {et~al.} 2009, Publications
  of the Astronomical Society of Australia, 26, 243

\bibitem[{{Herwig}(2005)}]{herwig2005}
{Herwig}, F. 2005, \araa, 43, 435

\bibitem[{{Hill} {et~al.}(2002){Hill}, {Plez}, {Cayrel}, {Beers},
  {Nordstr{\"o}m}, {Andersen}, {Spite}, {Spite}, {Barbuy}, {Bonifacio},
  {Depagne}, {Fran{\c c}ois}, \& {Primas}}]{hill2002}
{Hill}, V., {Plez}, B., {Cayrel}, R., {et~al.} 2002, \aap, 387, 560

\bibitem[{{Hirschi}(2007)}]{hirschi2007}
{Hirschi}, R. 2007, \aap, 461, 571

\bibitem[{{Hirschi} \& {et al.}(2006)}]{hirschi2006}
{Hirschi}, R., \& {et al.} 2006, in Reviews in Modern Astronomy, Vol.~19,
  Reviews in Modern Astronomy, ed. S.~{Roeser}, 101

\bibitem[{{Ito} {et~al.}(2013){Ito}, {Aoki}, {Beers}, {Tominaga}, {Honda}, \&
  {Carollo}}]{ito2013}
{Ito}, H., {Aoki}, W., {Beers}, T.~C., {et~al.} 2013, \apj, 773, 33

\bibitem[{{Ito} {et~al.}(2009){Ito}, {Aoki}, {Honda}, \& {Beers}}]{ito2009}
{Ito}, H., {Aoki}, W., {Honda}, S., \& {Beers}, T.~C. 2009, \apjl, 698, L37

\bibitem[{{K{\"a}ppeler} {et~al.}(2011){K{\"a}ppeler}, {Gallino}, {Bisterzo},
  \& {Aoki}}]{kappeler2011}
{K{\"a}ppeler}, F., {Gallino}, R., {Bisterzo}, S., \& {Aoki}, W. 2011, Reviews
  of Modern Physics, 83, 157

\bibitem[{{Keller} {et~al.}(2014){Keller}, {Bessell}, {Frebel}, {Casey},
  {Asplund}, {Jacobson}, {Lind}, {Norris}, {Yong}, {Heger}, {Magic}, {da
  Costa}, {Schmidt}, \& {Tisserand}}]{keller2014}
{Keller}, S.~C., {Bessell}, M.~S., {Frebel}, A., {et~al.} 2014, \nat, 506, 463

\bibitem[{{Kupka} {et~al.}(2000){Kupka}, {Ryabchikova}, {Piskunov}, {Stempels},
  \& {Weiss}}]{kupka2000}
{Kupka}, F.~G., {Ryabchikova}, T.~A., {Piskunov}, N.~E., {Stempels}, H.~C., \&
  {Weiss}, W.~W. 2000, Baltic Astronomy, 9, 590

\bibitem[{{Kurucz}(1995)}]{kurucz1995}
{Kurucz}, R.~L. 1995, in Astronomical Society of the Pacific Conference Series,
  Vol.~78, Astrophysical Applications of Powerful New Databases, ed. S.~J.
  {Adelman} \& W.~L. {Wiese}, 205

\bibitem[{{Lee} {et~al.}(2013){Lee}, {Beers}, {Masseron}, {Plez}, {Rockosi},
  {Sobeck}, {Yanny}, {Lucatello}, {Sivarani}, {Placco}, \& {Carollo}}]{lee2013}
{Lee}, Y.~S., {Beers}, T.~C., {Masseron}, T., {et~al.} 2013, \aj, 146, 132

\bibitem[{{Lind} {et~al.}(2012){Lind}, {Bergemann}, \& {Asplund}}]{lind2012}
{Lind}, K., {Bergemann}, M., \& {Asplund}, M. 2012, \mnras, 427, 50

\bibitem[{{Lucatello} {et~al.}(2003){Lucatello}, {Gratton}, {Cohen}, {Beers},
  {Christlieb}, {Carretta}, \& {Ram{\'{\i}}rez}}]{lucatello2003}
{Lucatello}, S., {Gratton}, R., {Cohen}, J.~G., {et~al.} 2003, \aj, 125, 875

\bibitem[{{Lucatello} {et~al.}(2005){Lucatello}, {Tsangarides}, {Beers},
  {Carretta}, {Gratton}, \& {Ryan}}]{lucatello2005}
{Lucatello}, S., {Tsangarides}, S., {Beers}, T.~C., {et~al.} 2005, \apj, 625,
  825

\bibitem[{{Masseron} {et~al.}(2012){Masseron}, {Johnson}, {Lucatello},
  {Karakas}, {Plez}, {Beers}, \& {Christlieb}}]{masseron2012}
{Masseron}, T., {Johnson}, J.~A., {Lucatello}, S., {et~al.} 2012, The
  Astrophysical Journal, 751, 14

\bibitem[{{Masseron} {et~al.}(2010){Masseron}, {Johnson}, {Plez}, {van Eck},
  {Primas}, {Goriely}, \& {Jorissen}}]{masseron2010}
{Masseron}, T., {Johnson}, J.~A., {Plez}, B., {et~al.} 2010, Astronomy and
  Astrophysics, 509, A93

\bibitem[{{McWilliam}(1998)}]{mcwilliam1998}
{McWilliam}, A. 1998, \aj, 115, 1640

\bibitem[{{Meynet} {et~al.}(2006){Meynet}, {Ekstr{\"o}m}, \&
  {Maeder}}]{meynet2006}
{Meynet}, G., {Ekstr{\"o}m}, S., \& {Maeder}, A. 2006, \aap, 447, 623

\bibitem[{{Meynet} {et~al.}(2010){Meynet}, {Hirschi}, {Ekstrom}, {Maeder},
  {Georgy}, {Eggenberger}, \& {Chiappini}}]{meynet2010}
{Meynet}, G., {Hirschi}, R., {Ekstrom}, S., {et~al.} 2010, Astronomy and
  Astrophysics, 521, A30

\bibitem[{{Nomoto} {et~al.}(2013){Nomoto}, {Kobayashi}, \&
  {Tominaga}}]{nomoto2013}
{Nomoto}, K., {Kobayashi}, C., \& {Tominaga}, N. 2013, \araa, 51, 457

\bibitem[{{Norris} {et~al.}(2007){Norris}, {Christlieb}, {Korn}, {Eriksson},
  {Bessell}, {Beers}, {Wisotzki}, \& {Reimers}}]{norris2007}
{Norris}, J.~E., {Christlieb}, N., {Korn}, A.~J., {et~al.} 2007, The
  Astrophysical Journal, 670, 774

\bibitem[{{Norris} {et~al.}(2013{\natexlab{a}}){Norris}, {Bessell}, {Yong},
  {Christlieb}, {Barklem}, {Asplund}, {Murphy}, {Beers}, {Frebel}, \&
  {Ryan}}]{norris2013a}
{Norris}, J.~E., {Bessell}, M.~S., {Yong}, D., {et~al.} 2013{\natexlab{a}},
  \apj, 762, 25

\bibitem[{{Norris} {et~al.}(2013{\natexlab{b}}){Norris}, {Yong}, {Bessell},
  {Christlieb}, {Asplund}, {Gilmore}, {Wyse}, {Beers}, {Barklem}, {Frebel}, \&
  {Ryan}}]{norris2013b}
{Norris}, J.~E., {Yong}, D., {Bessell}, M.~S., {et~al.} 2013{\natexlab{b}},
  \apj, 762, 28

\bibitem[{{Piau} {et~al.}(2006){Piau}, {Beers}, {Balsara}, {Sivarani},
  {Truran}, \& {Ferguson}}]{piau2006}
{Piau}, L., {Beers}, T.~C., {Balsara}, D.~S., {et~al.} 2006, \apj, 653, 300

\bibitem[{{Pignatari} {et~al.}(2010){Pignatari}, {Gallino}, {Heil}, {Wiescher},
  {K{\"a}ppeler}, {Herwig}, \& {Bisterzo}}]{pignatari2010}
{Pignatari}, M., {Gallino}, R., {Heil}, M., {et~al.} 2010, \apj, 710, 1557

\bibitem[{{Pignatari} {et~al.}(2008){Pignatari}, {Gallino}, {Meynet},
  {Hirschi}, {Herwig}, \& {Wiescher}}]{pignatari2008}
{Pignatari}, M., {Gallino}, R., {Meynet}, G., {et~al.} 2008, The Astrophysical
  Journal, 687, L95

\bibitem[{{Placco} {et~al.}(2014){Placco}, {Frebel}, {Beers}, {Christlieb},
  {Lee}, {Kennedy}, {Rossi}, \& {Santucci}}]{placco2014}
{Placco}, V.~M., {Frebel}, A., {Beers}, T.~C., {et~al.} 2014, \apj, 781, 40

\bibitem[{{Placco} {et~al.}(2013){Placco}, {Frebel}, {Beers}, {Karakas},
  {Kennedy}, {Rossi}, {Christlieb}, \& {Stancliffe}}]{placco2013}
---. 2013, \apj, 770, 104

\bibitem[{{Placco} {et~al.}(2010){Placco}, {Kennedy}, {Rossi}, {Beers}, {Lee},
  {Christlieb}, {Sivarani}, {Reimers}, \& {Wisotzki}}]{placco2010}
{Placco}, V.~M., {Kennedy}, C.~R., {Rossi}, S., {et~al.} 2010, \aj, 139, 1051

\bibitem[{{Placco} {et~al.}(2011){Placco}, {Kennedy}, {Beers}, {Christlieb},
  {Rossi}, {Sivarani}, {Lee}, {Reimers}, \& {Wisotzki}}]{placco2011}
{Placco}, V.~M., {Kennedy}, C.~R., {Beers}, T.~C., {et~al.} 2011, \aj, 142, 188

\bibitem[{{Reddy} {et~al.}(2003){Reddy}, {Tomkin}, {Lambert}, \& {Allende
  Prieto}}]{reddy2003}
{Reddy}, B.~E., {Tomkin}, J., {Lambert}, D.~L., \& {Allende Prieto}, C. 2003,
  Monthly Notice of the Royal Astronomical Society, 340, 304

\bibitem[{{Roederer}(2013)}]{roederer2013}
{Roederer}, I.~U. 2013, The Astronomical Journal, 145, 26

\bibitem[{{Ryan} {et~al.}(1999){Ryan}, {Norris}, \& {Beers}}]{ryan1999}
{Ryan}, S.~G., {Norris}, J.~E., \& {Beers}, T.~C. 1999, \apj, 523, 654

\bibitem[{{Sackmann} \& {Boothroyd}(1992)}]{sackmann1992}
{Sackmann}, I.-J., \& {Boothroyd}, A.~I. 1992, The Astrophysical Journal
  Letters, 392, L71

\bibitem[{{Sbordone} {et~al.}(2010){Sbordone}, {Bonifacio}, {Caffau}, {Ludwig},
  {Behara}, {Gonz{\'a}lez Hern{\'a}ndez}, {Steffen}, {Cayrel}, {Freytag},
  {van't Veer}, {Molaro}, {Plez}, {Sivarani}, {Spite}, {Spite}, {Beers},
  {Christlieb}, {Fran{\c c}ois}, \& {Hill}}]{sbordone2010}
{Sbordone}, L., {Bonifacio}, P., {Caffau}, E., {et~al.} 2010, Astronomy and
  Astrophysics, 522, A26

\bibitem[{{Sivarani} {et~al.}(2006){Sivarani}, {Beers}, {Bonifacio}, {Molaro},
  {Cayrel}, {Herwig}, {Spite}, {Spite}, {Plez}, {Andersen}, {Barbuy},
  {Depagne}, {Hill}, {Fran{\c c}ois}, {Nordstr{\"o}m}, \&
  {Primas}}]{sivarani2006}
{Sivarani}, T., {Beers}, T.~C., {Bonifacio}, P., {et~al.} 2006, \aap, 459, 125

\bibitem[{{Sneden}(1973)}]{sneden1973}
{Sneden}, C. 1973, The Astrophysical Journal, 184, 839

\bibitem[{{Sneden} {et~al.}(2008){Sneden}, {Cowan}, \& {Gallino}}]{sneden2008}
{Sneden}, C., {Cowan}, J.~J., \& {Gallino}, R. 2008, \araa, 46, 241

\bibitem[{{Sneden} {et~al.}(2003){Sneden}, {Cowan}, {Lawler}, {Ivans},
  {Burles}, {Beers}, {Primas}, {Hill}, {Truran}, {Fuller}, {Pfeiffer}, \&
  {Kratz}}]{sneden2003}
{Sneden}, C., {Cowan}, J.~J., {Lawler}, J.~E., {et~al.} 2003, \apj, 591, 936

\bibitem[{{Sobeck} {et~al.}(2011){Sobeck}, {Kraft}, {Sneden}, {Preston},
  {Cowan}, {Smith}, {Thompson}, {Shectman}, \& {Burley}}]{sobeck2011}
{Sobeck}, J.~S., {Kraft}, R.~P., {Sneden}, C., {et~al.} 2011, The Astronomical
  Journal, 141, 175

\bibitem[{{Spite} {et~al.}(2013){Spite}, {Caffau}, {Bonifacio}, {Spite},
  {Ludwig}, {Plez}, \& {Christlieb}}]{spite2013}
{Spite}, M., {Caffau}, E., {Bonifacio}, P., {et~al.} 2013, Astronomy and
  Astrophysics, 552, A107

\bibitem[{{Stancliffe}(2009)}]{stancliffe2009}
{Stancliffe}, R.~J. 2009, \mnras, 394, 1051

\bibitem[{{Stancliffe} {et~al.}(2007){Stancliffe}, {Glebbeek}, {Izzard}, \&
  {Pols}}]{stancliffe2007}
{Stancliffe}, R.~J., {Glebbeek}, E., {Izzard}, R.~G., \& {Pols}, O.~R. 2007,
  \aap, 464, L57

\bibitem[{{Starkenburg} {et~al.}(2014){Starkenburg}, {Shetrone}, {McConnachie},
  \& {Venn}}]{starkenburg2014}
{Starkenburg}, E., {Shetrone}, M.~D., {McConnachie}, A.~W., \& {Venn}, K.~A.
  2014, ArXiv e-prints, arXiv:1404.0385

\bibitem[{{Suda} {et~al.}(2004){Suda}, {Aikawa}, {Machida}, {Fujimoto}, \&
  {Iben}}]{suda2004}
{Suda}, T., {Aikawa}, M., {Machida}, M.~N., {Fujimoto}, M.~Y., \& {Iben}, Jr.,
  I. 2004, The Astrophysical Journal, 611, 476

\bibitem[{{Talon} \& {Charbonnel}(2005)}]{talon2005}
{Talon}, S., \& {Charbonnel}, C. 2005, \aap, 440, 981

\bibitem[{{Thielemann} {et~al.}(2011){Thielemann}, {Arcones}, {K{\"a}ppeli},
  {Liebend{\"o}rfer}, {Rauscher}, {Winteler}, {Fr{\"o}hlich}, {Dillmann},
  {Fischer}, {Martinez-Pinedo}, {Langanke}, {Farouqi}, {Kratz}, {Panov}, \&
  {Korneev}}]{thielemann2011}
{Thielemann}, F.-K., {Arcones}, A., {K{\"a}ppeli}, R., {et~al.} 2011, Progress
  in Particle and Nuclear Physics, 66, 346

\bibitem[{{Tominaga} {et~al.}(2013){Tominaga}, {Iwamoto}, \&
  {Nomoto}}]{tominaga2013}
{Tominaga}, N., {Iwamoto}, N., \& {Nomoto}, K. 2013, arXiv:astro-ph/1309.6734

\bibitem[{{Tominaga} {et~al.}(2007){Tominaga}, {Umeda}, \&
  {Nomoto}}]{tominaga2007}
{Tominaga}, N., {Umeda}, H., \& {Nomoto}, K. 2007, \apj, 660, 516

\bibitem[{{Truran}(1981)}]{truran1981}
{Truran}, J.~W. 1981, \aap, 97, 391

\bibitem[{{Umeda} \& {Nomoto}(2003)}]{umeda2003}
{Umeda}, H., \& {Nomoto}, K. 2003, Nature, 422, 871

\bibitem[{{Umeda} \& {Nomoto}(2005)}]{umeda2005}
---. 2005, \apj, 619, 427

\bibitem[{{Yong} {et~al.}(2013){Yong}, {Norris}, {Bessell}, {Christlieb},
  {Asplund}, {Beers}, {Barklem}, {Frebel}, \& {Ryan}}]{yong2013}
{Yong}, D., {Norris}, J.~E., {Bessell}, M.~S., {et~al.} 2013, The Astrophysical
  Journal, 762, 26

\end{thebibliography}

\clearpage

\end{document}